# Organometallic-Inorganic Hybrid MXenes with Tunable Superconductivity

Qi Fan[1,2]†, Tao Bo[1,3]†, Wei Guo[4], Minghua Chen[1], Qing Tang[1], Yicong Yang[1], Mian Li[1,3], Ke Chen[1,3], Fangfang Ge[1,3], Jialu Li[5], Sicong Qiao[6], Changda Wang[6], Li Song[6], Lijing Yu[7], Jinghua Guo[5], Michael Naguib[8], Zhifang Chai[1,3], Qing Huang[1,3], Chaochao Dun[5]*, Ning Kang[4]*, Yury Gogotsi[9]*, Kun Liang[1,3]*

**Affiliations:**

[1]Zhejiang Key Laboratory of Data-Driven High-Safety Energy Materials and Applications Ningbo Key Laboratory of Special Energy Materials and Chemistry, Ningbo Institute of Materials Technology and Engineering, Chinese Academy of Sciences; Ningbo 315201, P.R. China.

[2]University of Chinese Academy of Sciences; 19 A Yuquan Rd, Shijingshan District, Beijing 100049, P. R. China.

[3]Qianwan Institute of CNITECH; Ningbo 315336, P.R. China.

[4]Key Laboratory for the Physics and Chemistry of Nanodevices and Department of Electronics, Peking University; Beijing 100871, P.R. China

[5]Lawrence Berkeley National Laboratory; Berkeley, CA 94720, USA.

[6]National Synchrotron Radiation Laboratory, Key Laboratory of Precision and Intelligent Chemistry, School of Nuclear Science and Technology, University of Science and Technology of China; Hefei 230029, P. R. China.

[7]Key Laboratory of Materials and Surface Technology (Ministry of Education), Sichuan Energy Equipment Intelligent Engineering Research Center, School of Materials Science and Engineering, Xihua University, Chengdu 610039, P. R. China.

[8]Department of Physics and Engineering Physics, Tulane University; New Orleans, Louisiana 70118, USA.

[9]Department of Materials Science and Engineering, A. J. Drexel Nanomaterials Institute, Drexel University; Philadelphia, PA 19104, USA.

†These authors contributed equally to this work.

*Corresponding author. Email: cdun@lbl.gov; nkang@pku.edu.cn; gogotsi@drexel.edu; kliang@nimte.ac.cn

## Abstract

Ti-based two-dimensional transition-metal carbides (MXenes) have attracted attention due to their superior properties and are being explored across various applications[1,2]. Despite their versatile properties, superconductivity has never been demonstrated, not even predicted, for this important group of 2D materials. In this work, we have introduced an electrochemical intercalation protocol to construct versatile organometallic-inorganic hybrid MXenes and achieved tunable superconductivity in the metallocene-modified layered crystals. Through structural editing of MXene matrix at atomic scale and meticulously modulated intercalation route, $Ti_3C_2T_x$ intercalated with metallocene species exhibits a superconductive transition temperature ($T_c$) of 10.2 K. Guest intercalation induced electron filling and strain engineering are responsible for the emerging superconductivity in this intrinsically non-superconducting material. Theoretically, simulated electron-phonon interaction effects further elucidate the nature of the changes in $T_c$. Furthermore, the $T_c$ of crafted artificial superlattices beyond Ti-based MXenes have been predicted, offering a general strategy for engineering superconductivity and magnetism in layered hybrid materials.

# Introduction

Layered van der Waals (vdW) materials, such as graphite, transition metal dichalcogenides, and transition metal carbides and/or nitrides (MXenes), offer attractive physical properties and various electronic structures[3,4]. The advances in the field of 2D materials contributed to a renewed interest in these layered compounds, whose weak interlayer coupling allows easy intercalation and preparation of single- or few-layered nanosheets[5]. vdW gaps between adjacent layers also open up vast possibilities for achieving continuous regulation of quantum phenomena by intercalating various species, forming intercalated compounds[6]. By incorporating electronic carriers with larger sizes, guest-host interactions enable structural editing and properties tailoring of host vdW materials based on the orbital occupation tuned with electron filling and the interlayer coupling regulated by decreasing dimensionality[7,8]. Notably, a combined effect between charge doping and strain engineering may perturb the subtle equilibrium of quantum states[9,10]. Among many intercalation protocols, electrochemical intercalation offers a unique and efficient platform with precise controllability over intercalation level, to manipulate the stoichiometry or structure of the materials to achieve targeted properties, such as superconductivity, magnetism, and ferroelectricity[7,11,12].

Superconductivity is a macroscopic quantum phenomenon with the electrons forming "Cooper pairs" and electric currents flowing without dissipation below a critical transition temperature ($T_c$). It is particularly important for electric transportation, nuclear magnetic diagnosis, and quantum computing[13,14]. The reduced dimension bestows 2D superconductors with significant discoveries, including quantum metallic state, Ising superconductivity, and continuous phase transition[15]. In contrast to 3D superconductors, 2D superconductors offer enhanced tunability for the desired superconductivity performance through adjusting sample thickness, modulating carrier density, constructing coupled heterostructures, introducing biaxial strain, etc[16-18]. For example, the superconductivity of some transition metal dichalcogenides (e.g., $WTe_2$ and $MoTe_2$) can be engineered by reducing their thickness to a monolayer for a greatly improved $T_c$[19]. However, its practical application has been plagued by processing challenges to meet the quality requirements and environmental stability[20,21]. Fortunately, intercalated vdW layered materials have tailored emergent novel quantum phenomena and unprecedented properties exceeding both bulk materials and monolayer samples while maintaining quality and stability[6,16,22]. 2D carbides and nitrides of transition metals known as MXenes have intrinsic metallicity and allow structure editing at the atomic scale. However, little has been done to explore their superconductive properties[1]. The first experimental report was about ultrathin α-$Mo_2C$ with $T_c$ of 3 K[23], and chemically tunable superconductivity in MXene was predicted by B. Yakobson et al.[24]. A transition behavior from semiconductor to superconductor was engineered via chemical transformations of surface functional groups in Nb-based MXenes with the highest $T_c$ of 7.1 K[25]. Similarly, the $NH_3$-treated $Nb_2CT_x$ displayed the anisotropic superconducting transitions produced by the high-temperature reaction[26]. Harnessing intercalation chemistry in MXenes contributes to favorable electron filling and allows strain engineering of each nanosheet, providing an opportunity to control superconductive properties[27]. While titanium-based MXenes are the most studied, no superconductivity has been reported in those materials. With their extreme strength and easy processability[1], these MXenes made of abundant elements could potentially provide low-cost flexible superconductor tapes and printed devices that cannot be made using conventional materials.

Herein, we report on an electrochemically modulated structural editing protocol in the organometallic-inorganic hybrid MXenes to achieve tunable superconductivity in Ti-based MXenes. The control of superconductivity via electrochemical intercalation includes a variation of intercalants, co-intercalated solvents, and editable MXenes hosts involving M/X sites, as well as the number of transition metal layers and surface terminations. Transport properties measurements and theoretical calculations indicate that varying degrees of electron transfer from cyclopentadienyls to the MXene lattice induces adjustable superconducting states, facilitated by tunable electron-phonon coupling effect from reduced dimensionality and strain engineering. This editing strategy could chart a versatile path to designing novel superconductors beyond Ti-based MXenes.

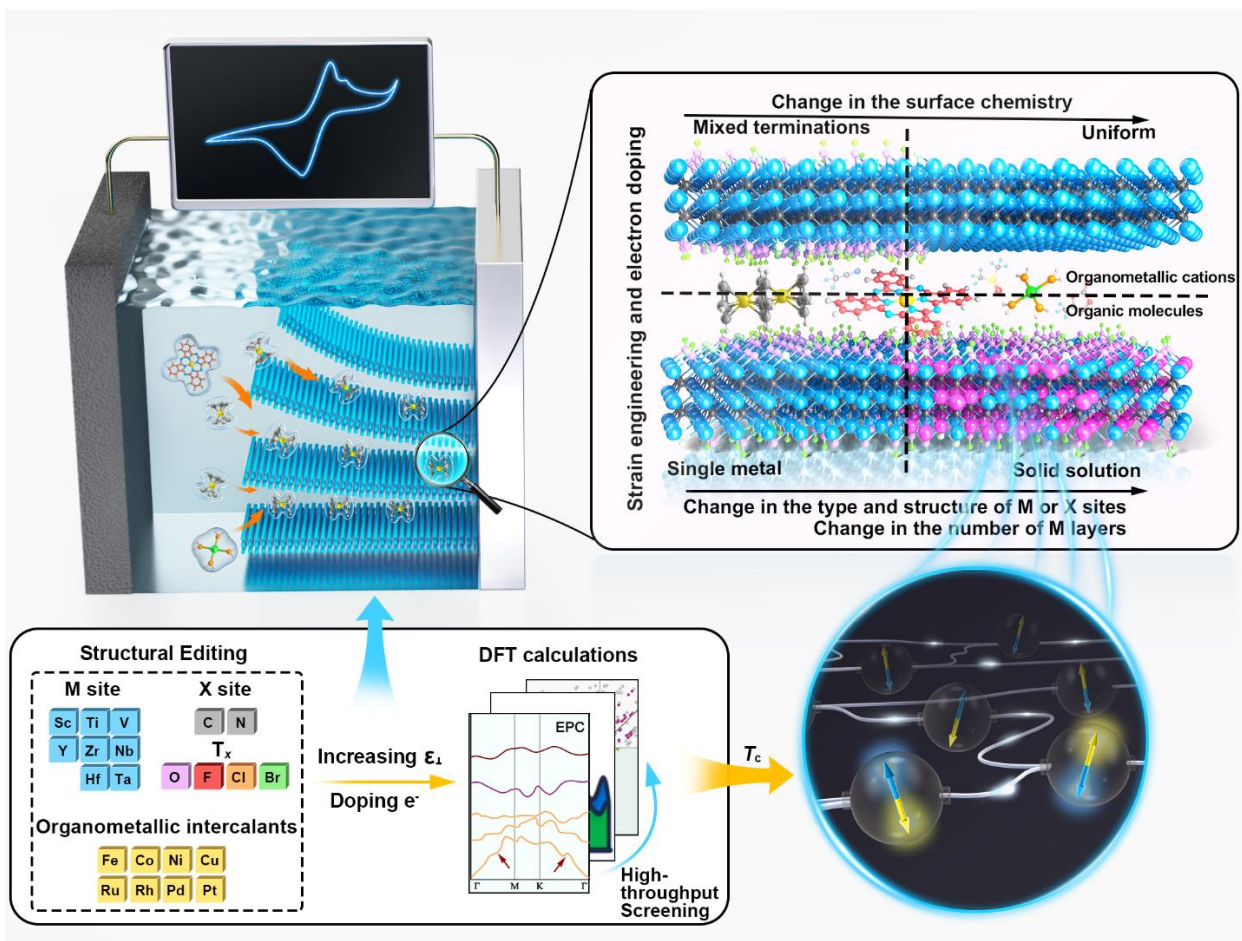

**Fig. 1 Schematic illustration of an electrochemically modulated structural editing protocol in the organometallic-inorganic hybrid MXenes structures to achieve tunable superconductivity in Ti-based MXenes.** The control of superconductive property via electrochemical intercalation includes variations of organometallic cations, co-intercalated organic molecules, and the editable host MXenes materials involving changes in the surface chemistry, structure, the type of M or X sites, and the number of M layers. This editing strategy produces diverse superconductors beyond Ti-based MXenes. Their superconducting critical temperatures can be deduced through DFT calculations and high-throughput screening.

## Organometallic-inorganic hybrid MXenes

Fig. 1 illustrates the construction of an organometallic-inorganic hybrid MXenes superlattice through the electrochemical intercalation of organometallic species as pillar molecules and charge donors. Diverse multi-layered MXenes (*m*-MXene) were synthesized via wet chemical etching and molten salt etching routes serving as the host materials (Supplementary Fig. 1). Coupled with a carbon rod, the electrochemical setup operates under constant current sweeping to continuously drive metallocene cations into the $Ti_3C_2T_x$ ($T_x$=-F/O/OH) *m*-MXene electrodes. As shown in Supplementary Fig. 2a, a prominent cathodic peak appears around -1.39 V (vs. carbon), indicating the intercalation of $[Co(Cp)_2]^+$ ions and subsequent reduction to $Co(Cp)_2$. A constant current of 2 mA was applied to ensure a steady and uniform electrochemical intercalation (Supplementary Fig. 2b), which completes within 5 h, yielding what we term as Co-*m*-MXene. The structural editing by electrochemical intercalation applies to a broad range of layered vdW materials, including $Nb_2CT_x$, $Ti_3CNT_x$, and $Ti_3C_2Cl_2$ MXenes and transition metal dichalcogenides (TMDs), such as $TiS_2$, $TaS_2$, and $SnSe_2$. As shown in Fig. 1, the structural diversity and adjustability of organometallic intercalants allow them to incorporate useful properties and opportunities into the versatile 2D material matrix. For example, introducing $Co(Cp)_2$ species may unlock extraordinary electronic and magnetic properties in Ti-based MXenes. Furthermore, organometallic compounds containing noble metals (e.g. Pt, Ru and Pd) are expected to provide efficient catalytic active sites and striking stability, under the protection of layered hosts. Our electrochemical intercalation model shows substantial efficiency compared to conventional intercalation method for TMDs materials, which typically involve protracted, week-long reactions[22].

The effective intercalation of guest species is corroborated by X-ray diffraction (XRD), high-angle annular dark-field scanning transmission electron microscopy (HAADF-STEM), and energy-dispersive X-ray spectroscopy (EDX) elemental mapping (Fig. 2a-e). XRD patterns (Fig. 2a) exhibit a notable shift in the (002) peak of $Ti_3C_2$-*m*-MXene, with the corresponding interlayer distance increasing from 0.98 nm to 1.57 nm. This shift aligns with the size of a single molecular layer of $Co(Cp)_2$ in the expanded vdW gap. Noteworthily, none of the broad (00*l*) reflections often observed in mixed polytype host MXene materials is present, ascribed to a superior layer stacking order[28]. It could be more accurately determined in the HAADF-STEM images by measuring the interlayer spacing in the sectional image of the intercalated compounds (Fig. 2b, c). Corresponding EDS mapping and electron energy loss spectroscopy (EELS) curve further support the uniform incorporation of $[Co(Cp)_2]$ species monolayers in the $Ti_3C_2$-MXene interlayer spaces (Fig. 2d and Supplementary Fig. 3). According to the EELS spectral information detected in the interlayer region, the valent state of cobalt in the intercalated compound is close to +2[29], and the infrared signal of cobaltocene can still be detected in the intercalated compounds (Supplementary Fig. 4). Furthermore, the orientation of cobaltocene among vdW gap of MXene warrants further clarification, which can be well proved by the one-dimensional (1D) electron density map along the *c*-axis of the Co-*m*-MXene hybrid, following a "side-on" pattern (Fig. 2e, f)[30,31]. Density functional theory (DFT) calculations (Supplementary Fig. 5) provide additional insight from the relative stability parameter, defined as the DFT total energy difference between two orientations of cobaltocene ($E_{rel} = E_{end-on} - E_{side-on}$)[32]. A positive value of $E_{rel}$ across all terminations confirms the side-on orientation is more stable. The formation of Co-*m*-$Nb_2CT_x$, Co-*m*-$Ti_3CNT_x$, $[Pt(NH_2)_4]^+$-*m*-MXene, $[Ru(NH_2)_6]^{3+}$-*m*-MXene, $[Pd(NH_2)_4]^{2+}$-*m*-MXene, CoPc-*m*-MXene and Co-*m*-$SnSe_2$ can also be verified by the SEM-EDS mapping and XRD results (Supplementary Fig. 6-9).

To further verify the conjecture and unveil the ions intercalation process, we employed Ti and Co K-edge X-ray absorption near edge structure (XANES) spectroscopy and X-ray photoelectron spectroscopy (XPS) for monitoring the chemical state with the addition of Co (Fig. 2g and Supplementary Fig. 10a, b). The electron filling effect is evidenced by a shift in the Ti absorption peak towards lower energy, indicating the decreased valence state of Ti from 2.92 to 2.75 (i.e., by≈0.17$\bar{e}$ per Ti atom and 0.51$\bar{e}$ per $Ti_3C_2T_x$) after the electrochemical intercalation and electron hopping from [Co(Cp)$_2$]. Concurrently, most [Co(Cp)$_2$]$^+$ cations are electrochemically reduced to [Co(Cp)$_2$] species within the vdW gaps of MXene, consistent with the observed results in EELS (Supplementary Fig. 3).

There are also distinct changes in the Raman spectra after intercalation, where the reduced layer-to-layer interaction or charge-doping from the metallocene cations results in the observable redshift of the $A_{1g\ (Ti,\ C,\ T_x)}$ mode and disappearance of the shear mode, approaching monolayer characteristics (Fig. 2h)[16]. Recent studies reported that in-plane strain can be examined through Raman spectroscopy[33,34]. After accurate peak fitting (Supplementary Fig. 11), a clear red shift of ~3.9 cm$^{-1}$ was observed. The intercalation-induced tensile strain (δ) on the z-axis could be approximatively calculated by $\Delta\omega = \delta*\omega/2.66$, in which ω is the reference frequency and Δω is the change of Raman peak position. A calculated strain of ~4.9% for the MXene basal plane is introduced after the intercalation of [Co(Cp)$_2$] species.

Atomic force microscopy (AFM) studies support the interlayer spacing expansion observed in the XRD studies (Supplementary Fig. 12). Mapping of the variation of Young's modulus, $E$, allowed us to visualize the intercalation effect in a spatially resolved manner. Three-dimensional nanoscale mapping of $E$ for the same MXene flake demonstrates an overall increase in $E$ across the whole flake, with distribution fitted by Gaussian profiles, indicating increased stiffness after intercalation (Fig. 2i and Supplementary Fig. 13). Geometric phase analysis (GPA) maps and corresponding high resolution TEM images (Fig. 2j and Supplementary Fig. 14) further revealed dislocation centers with interface strains that appeared between intercalants and MXene hosts, which are theoretically conducive to electron transfer[35]. The Ti-Ti3 interatomic distance ($d_{Ti-Ti3}$) was enlarged from 513.7 pm to 540.4 pm in the Co-*m*-MXene, ascribed to lattice strains induced by intercalation (Supplementary Fig. 15). Using the distance between the (111) planes of bulk TiC for reference, the variation of out-of-plane strain ($\varepsilon_\perp$) was calculated to be ~5.3% after intercalation[25]. Consequently, we have a clearer depiction of the structural changes, charge transfer, and interaction within the intercalated Co-*m*-MXene compound (Fig. 2f).

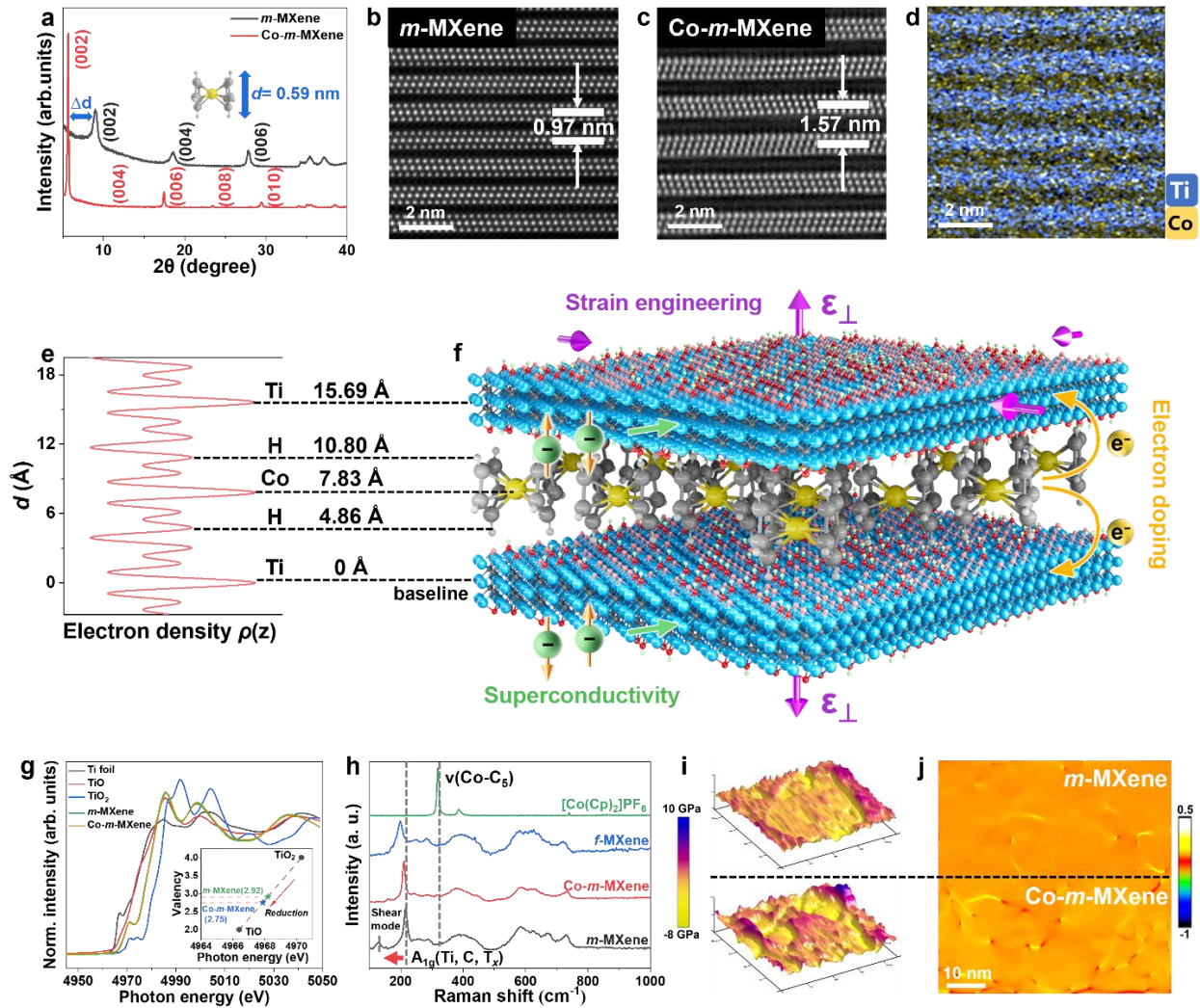

**Fig. 2 Strain engineering and electron doping induced by intercalated [Co(Cp)$_2$]. a,** XRD patterns of *m*-MXene and Co-*m*-MXene. Intercalation of [Co(Cp)$_2$] species results in a distinctly left shift of (002) peak of MXene, and Δd is consistent with the size of intercalants. (**b** and **c**) STEM images of *m*-MXene (**b**) and Co-*m*-MXene (**c**). All STEM images were acquired along the [11$\bar{2}$0] zone axis of MXenes. **d**, EDS mapping of Co-*m*-MXene representing [Co(Cp)$_2$] species existent between Ti$_3$C$_2$T$_x$ MXene layers. **e**, 1D electron density map along the c-axis of Co-*m*-MXene, providing structural information about the atomic arrangement and intercalant orientation. **f**, Schematic illustrations of strain engineering and electron doping from intercalated [Co(Cp)$_2$] for the MXene hosts, contributing to the emergent superconductivity. **g**, Ti K-edge XANES spectra of Ti$_3$C$_2$-*m*-MXene and Co-*m*-MXene. Inset: Average Ti oxidation state determination using the Ti K-edge energy shift of TiO and TiO$_2$ references. **h**, Raman spectra of *m*-MXene, Co-*m*-MXene, *f*-MXene, and [Co(Cp)$_2$]PF$_6$. **i**, Three-dimensional topography of the same MXene nanoflake before intercalation (top) and after intercalation (bottom). A deepened color in the latter indicates the enhanced Young's modulus and, thus, lattice strain. **j**, Strain maps of *m*-MXene (top) and Co-*m*-MXene (bottom). The points with reversed color indicate local pinning of layers and interface strains.

## Superconductivity in intercalated Ti-based MXenes

To date, there are no reports of the intrinsic superconductivity in Ti-based MXenes, primarily because titanium metal exhibits an ultralow superconductive transition temperature of 0.39 K, which contrasts sharply with 9.25 K for niobium metal[36]. We investigated the electronic transport and magnetic susceptibility of organometallic-inorganic Co-$m$-MXene hybrid in which MXene nanosheets were modified by electron filling and strain engineering. Fig. 3a shows the temperature dependence of the four-point probe resistivity of the Co-$m$-MXene superlattice from 300 to 2 K at zero magnetic field (Supplementary Fig. 16). The resistivity of this intercalated compound decreases continuously upon lowering the temperature, exhibiting a metallic behavior. Further cooling causes an abrupt drop of resistivity by several orders of magnitude to zero, indicating that the hybrids undergo a transition from a metallic to a superconducting state at an onset critical temperature of $T_{c\text{-onset}}$~8.3 K and a zero resistance temperature of $T_{c\text{-zero}}$~5.6 K, where $T_{c\text{-onset}}$ is the intersection point between the linear extrapolation of the normal state and the superconducting transition and $T_{c\text{-zero}}$ is the zero-resistance temperature. Unless further specified, the $T_c$ pertains to the $T_{c\text{-onset}}$[37]. The superconductivity of the superlattices was investigated by measuring the temperature dependencies of the resistivity and magnetization. As shown in Fig. 3b, $T_c$ gradually shifts to a lower temperature with increasing magnetic fields while the transition feature is suppressed and becomes less sharp. To estimate the upper critical field $H_{c2}(0)$, linear fitting of $T_c$ at different magnetic fields was done and extrapolated to the $H_{c2}(0)$ of 7.6 T, less than Bardeen-Cooper-Schrieffer (BSC) weak-coupling Pauli limit (Supplementary Fig. 17a). The superconducting characteristics of Co-$m$-MXene superlattice nanoflakes are also confirmed by measuring the resistance as a function of the magnetic field for different temperatures (Supplementary Fig. 17b). A critical field of 4 T is needed to suppress the superconductivity at 2 K. The isothermal magnetization curve (M-H curve) exhibits a typical diamagnetic character of superconductivity, known as the Meissner effect (Fig. 3c). This was further confirmed by the temperature dependence of magnetization (M-T curve) from 300 K to 2 K for both the zero-field-cooling (ZFC) and the field-cooling (FC) process. The magnetization of both curves sharply declined around 5.81 K, allowing the estimation of a lower bound for the superconducting volume fraction of Co-$m$-MXene as approximately 64.4 %, based on ZFC data at 2 K (Fig. 3d). In contrast, the pristine $m$-MXene shows a dramatically elevated resistivity and lacks any superconducting transition (Supplementary Fig. 18), in agreement with the published data[2].

By systematically varying intercalants, co-intercalated solvents and editing host MXene structures (including surface terminations, M and X elements, and the number of metal layers), we demonstrated the tunability of superconductivity in MXenes. Modifying M/X-site or surface terminations can alter the density of states (DOS), shift the Fermi level, and influence electron-phonon coupling (EPC), affecting the electronic behavior of MXenes SEM-EDS and XRD analyses verified the synthesis of new organometallic-inorganic Ti-based MXene hybrids (Fig. 3e, and Supplementary Fig. 19, 20). As shown in Fig. 3f, replacing $M_3C_2$ layers with $M_2C$ layers results in a higher superconducting transition temperature of $T_c$= 9.5 K for $Ti_2CT_x$. This can be attributed to enhanced electron-phonon interaction within the in-plane MXene flakes and interlayer coupling out-of-plane. When some Ti atoms were substituted with Nb atoms to form M'M"C (M' and M" representing different transition metals) solid solution, $TiNbCT_x$, the $T_c$ slightly decreased to 8.2 K. The thicker layers of out-of-plane ordered $Ti_2Mo_2C_3T_x$ had a lower $T_c$ of 6.4 K, primarily due to impeded interlayer coupling and enlarged band gap[38]. A mixture of C and N atoms at X sites also negatively impacted the charge carrier density, resulting in a lower $T_c$ of 6.4 K for $T_3CNT_x$ (Fig. 3g). Surface terminations of MXene mediate charge transfer between hosts with guest intercalants, which is revealed by the trend of $T_c$(-$T_x$) > $T_c$(-Cl) > $T_c$(-Br) (Fig. 3h). The coupling of MXene vibrational modes with functional groups suggests that the highest $T_c$ for $Ti_3C_2T_x$ with predominantly -F and -O terminations could be attributed to enhanced charge transfer or phonon dispersion and, thus, electron-phonon coupling. Introducing different organometallic donor molecules into the interlayer region regulates the electronic properties of MXene hybrids. The stable presence and uniform distribution of Fe(Cp)$_2$ was verified by the EDS-mapping of Fe-$m$-MXene (Supplementary Fig. 21). The stronger electron-donating capability of intercalated ferrocene species contributes to a notably higher $T_c$ of 10.2 K (Fig. 3i), which correlates with the lower valence state of Ti, as shown in Supplementary Fig. 22. Along with a larger lattice strain based on GPA analysis (Supplementary Fig. 21a-c), HAADF-STEM results show that the intercalated Fe(Cp)$_2$ induced approximately 10% lattice strain, which strongly influenced lattice vibration (Supplementary Fig. 21d). Similarly, co-intercalation of different organic solvents (e.g., dimethyl sulfoxide, propylene carbonate, and acetonitrile) offers an additional route to controlling interlayer conduction (Supplementary Fig. 23). The magnetic susceptibility measurements demonstrated the Meissner effect in these superconductive samples (Supplementary Fig. 24).

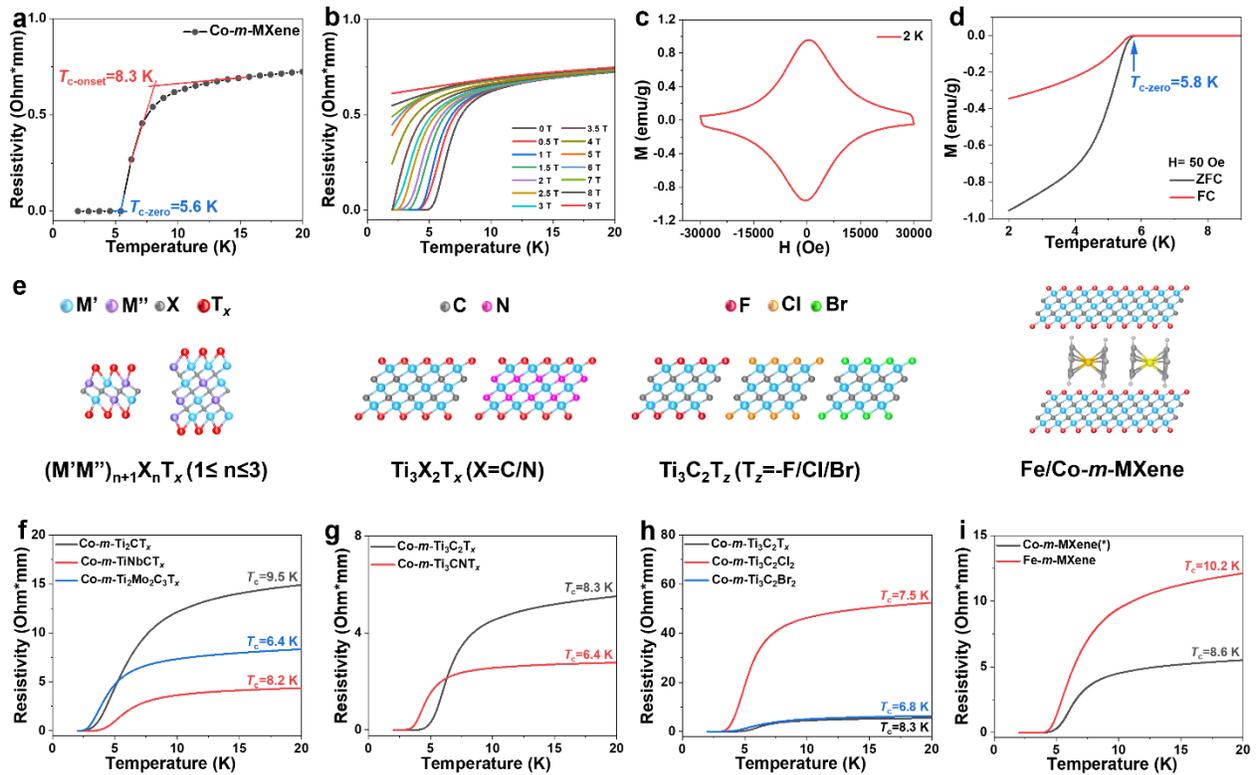

**Fig. 3 Tunable superconductivity in organometallic-inorganic Ti-based MXenes. a**, Temperature-dependent resistivity at zero magnetic field for the cold-pressed pellets of Co-$m$-MXene. **b**, Temperature dependence of the resistivity of Co-$m$-MXene under different magnetic fields. **c**, The isothermal magnetization curve (M-H curve) at 2 K showing the Meissner effect in Co-$m$-MXene. **d**, Temperature dependence of the magnetization of the Co-$m$-MXene superlattice under a magnetic field of 50 Oe for ZFC and FC processes. ZFC and FC correspond to the zero-field cooled and field cooled measurements, respectively. **e**, Crystal structure schematics of (M'M")$_{n+1}$X$_n$T$_x$ (1≤ n≤3), Ti$_3$X$_2$T$_x$ (X=C/N), Ti$_3$C$_2$T$_z$ (T$_z$= -F/Cl/Br) and Co/Fe-$m$-MXene. (**f** to **i**) Temperature-dependent resistivity for structure-edited Ti-based MXenes with various

organometallic intercalants: Co-*m*-Ti$_2$CT$_x$, Co-*m*-TiNbCT$_x$, and Co-*m*-Ti$_2$Mo$_2$C$_3$T$_x$ (**f**), Co-*m*-Ti$_3$C$_2$T$_x$ and Co-*m*-Ti$_3$CNT$_x$ (**g**), Co-*m*-Ti$_3$C$_2$T$_x$, Co-*m*-Ti$_3$C$_2$Cl$_2$ and Co-*m*-Ti$_3$C$_2$Br$_2$ (**h**), Co-*m*-MXene (\*) and Fe-*m*-MXene (**i**). Co-*m*-MXene (\*) represents its difference from Co-*m*-MXene, as its electrochemical intercalation was conducted under a current of 0.5 mA.

## Origin of tunable superconductivity in organometallic-inorganic hybrid MXenes

To understand the combined effect of strain engineering and charge doping on achieving variable $T_c$ values, we conducted resonant inelastic X-ray scattering (RIXS) characterization accompanied by first-principles DFT calculations. RIXS is an effective technique for probing the electronic states and electron correlations in Ti compounds. The Ti L$_3$-edge X-ray absorption spectroscopy (XAS) spectra (Fig. 4a) of *m*-MXene and Co-*m*-MXene display two prominent features in the energy ranges 455-461 eV and 461.2-468 eV, corresponding to L$_3$(2p3/2→3d) and L$_2$ (2p1/2→3d) absorptions respectively, while the former comprises Ti 3d-t$_{2g}$ state and 3d-e$_g$ state. The e$_g$ band is highly sensitive to changes in local environment; thus, its larger integrated area manifests an increased density of states for Ti upon electron doping from intercalated Co(Cp)$_2$[39,40]. Fig. 4b shows resonant photoemission maps at the Ti L$_{3,2}$ edge of Co-*m*-MXene, highlighting low energy loss (LEL) features like the elastic line (0-1 eV), d-d excitations (1-3 eV), and charge transfer (CT) beyond 4 eV. The low energy loss features are intimately associated with substantial EPC, and their intensity in Co-*m*-MXene exceeds the pristine sample at the t$_{2g}$ resonance (Fig. 4c and Supplementary Fig. 25). The enhanced intensity of energy loss after intercalation implies that more electrons are distributed in the t$_{2g}$ band, while a decreased excited feature at ~3.3 eV of the *d-d* transition and the e$_g$ resonance represent strong electron-electron correlation. These results suggest that the increase in the density of states of Ti 3d electrons at the Fermi level is responsible for the emergence of superconductivity. Next, we investigated the effects of electron/hole doping and tensile strain on the electronic structure through first-principles calculations. As shown in Supplementary Fig. 26, the differential charge density map of hybrid Co-*m*-MXene superlattice further clarifies the details of charge transfer at the heterointerfaces. Thus, Ti$_3$C$_2$T$_x$ (T$_x$ = -F or -O) receives electrons from the [Co(Cp)$_2$] layer, establishing some chemical linkage between them[41]. In agreement with the experimental results, a decreased charge transfer efficacy is predicted in the sequence of -F, -Cl, and -Br terminated MXene (Fig. 3g), while [Fe(Cp)$_2$] species render a larger degree of electron doping and, thus, a higher $T_c$ (Fig. 3h). The changes in superconductive parameters of electronic DOS at the Fermi level (N(E$_F$)), magnitude of the EPC ($\lambda_{qv}$), and the resulting $T_c$ are summarized in Fig. 4d. Applying tensile strain (enlarging Ti-Ti3 interatomic distances) and electron doping observably increase N(E$_F$) and $\lambda_{qv}$, leading to a higher $T_c$ value. A similar trend was also observed for Co-*m*-Ti$_2$CT$_x$ and Ti$_3$C$_2$T$_z$ (T$_z$=F/Cl/Br) (Supplementary Fig. 27, 28). To examine superconductivity in Co-*m*-MXene compounds, we derived phonon frequencies and EPC using density-functional perturbation theory and estimated critical transition temperatures through BCS theory and the McMillan equation. Unlike the limited EPC in the total Brillouin zone (BZ) of pristine MXene, 5% tensile strain and 0.5e doping from intercalation enhanced Ti$_z$, Ti$_{xy}$ and F$_z$ vibrations with softening phonons (marked by red arrows in Fig. 4e, f and Supplementary Fig. 29), facilitating the overall electron-phonon interactions. These vibrations dominate at the low frequencies and contribute 84% of the total EPC ($\lambda$= 0.475) to generate a $T_c$ of 7.63 K (Fig. 4g, h), consistent with the experimental observations (Fig. 3a). This synergistic effect can be further derived from the EPC integrated over all phonon branches and distributed in the BZ (Fig. 4i), where intense EPC appears around K point and along Γ-M or Γ-K lines and extends throughout the full BZ.

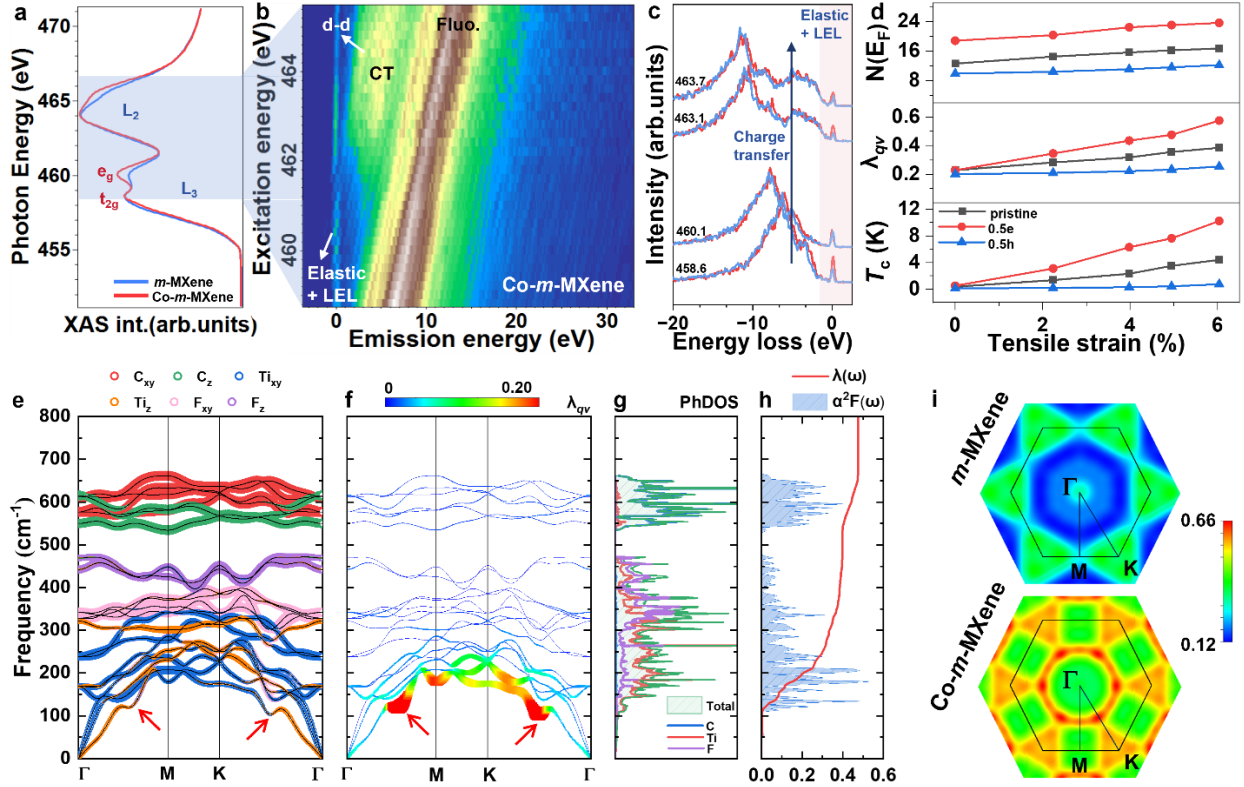

**Fig. 4 Mechanism of the emerging superconductivity in Co-*m*-MXene. a**, The XAS spectrum at the Ti $L_3$ edge of *m*-MXene and Co-*m*-MXene. The shaded regions indicate the incident X-ray energy (*hν*) ranges used in the RIXS maps. **b**, The corresponding RIXS map of Co-*m*-MXene, suggesting the existence of elastic emission and LEL at 0-1 eV with $Ti^{3+}$ (fluorescence and d-d excitations) at 1-4 eV in the sample. **c**, Ti $L_3$ RIXS spectra of *m*-MXene (blue line) and Co-*m*-MXene (red line) at the *hν* of 458.6 eV, 460.1 eV, 463.1 eV and 463.7 eV. **d**, $N(E_F)$, $\lambda_{qv}$, and $T_c$ for the Co-*m*-MXene after applying various tensile strains in a pristine state and after doping with 0.5 electrons and 0.5 holes. **e**, Phonon dispersions weighted by the motion modes of C, Ti and F atoms. The red, green, blue, orange, pink and purple colors represent $C_{xy}$, $C_z$, $Ti_{xy}$, $Ti_z$, $F_{xy}$ and $F_z$ modes, respectively. **f**, Phonon dispersions weighted by the magnitude of EPC $\lambda_{qv}$ for Co-*m*-MXene. **g**, The corresponding phonon density of states (PhDOS). **h**, The corresponding Eliashberg spectral function $\alpha^2F(\omega)$ and cumulative frequency-dependence of the EPC constant $\lambda(\omega)$. **i**, The integrated EPC distributions in the BZs for *m*-MXene (top) and Co-*m*-MXene (bottom).

## Conclusion

The entire family of MXenes without terminations is metallic[42]. Therefore, they all possess *a priori* host collective electron states that may enable exotic superconductivity. Our theoretical calculations have expanded the scope beyond Ti-based MXenes, identifying superconductors like $Sc_2C$, $Y_2C$, $V_2C$, $Hf_2C$, $Ta_2C$, $Zr_3C_2$, and others by electron doping and strain engineering from intercalated organometallic species (Supplementary Fig. 30, 31 and Supplementary Table 1-5). In conclusion, we demonstrate that electrochemically modulated interlayer engineering in the organometallic-inorganic MXene hybrid superlattices leads to a new electronic system with the $T_c$ of 10.2 K and possibly higher, originated from interlayered interactions of non-superconducting building blocks. The endless portfolio of organometallic intercalants and layered vdW material hosts offers numerous opportunities for tuning electronic, magnetic, and topological properties through precise tailoring of the Fermi level and spin-orbital effects in artificial superlattices. Establishing a database with high-throughput computational screening would facilitate the development and applications of those new superconductive materials.

## Methods

**Materials**

Ti$_3$AlC$_2$ (400 mesh, 99.99 %), Ti$_2$AlC (400 mesh, 99 %), TiNbAlC (400 mesh, 99 %), Ti$_2$Mo$_2$AlC (400 mesh, 99 %), Ti$_3$AlCN (400 mesh, 99 %), Nb$_2$AlC (400 mesh, 99 %) powders were purchased from Jilin 11 Technology Co., Ltd. SnSe$_2$ were purchased from Alfa Metal Materials Co., Ltd. HF solution (49 wt%), lithium chloride (LiCl), CdCl$_2$ (anhydrous, 99.99%), CdBr$_2$ (anhydrous, 99.99%), KCl (anhydrous, 99.99%), NaCl (anhydrous, 99.99%), lithium bis(trifluoromethylsulfonyl)amine (LiTFSI, 99.99%), bis(cyclopentadienyl)cobalt(III) hexafluorophosphate ([Co(Cp)$_2$]PF$_6$, (ferrocenylmethyl)-trimethylammonium chloride (FcNCl), dimethyl sulfoxide (DMSO, 99%), propylene carbonate (PC, 99.99%), and acetonitrile (ACN, 99.9%) were purchased from Aladdin Reagent Co. Ltd, China. Cobalt phthalocyanine (CoPc, 98%), hexaammineruthenium (III) chloride ([Ru(NH$_2$)$_6$]Cl$_3$, 98%), and tetraammineplatinum dinitrate ([Pt(NH$_2$)$_4$](NO$_3$)$_2$, 99.9%) were provided by Adamas-Beta. Hydrochloric acid (HCl) and hydrobromic acid (HBr) were purchased from Sinopharm Chemical Reagent Co., Ltd, China. High-purity water with a resistivity of 18.25 MΩ cm prepared from a UPH-IV ultrapure water purifier (China) was used throughout the experimental process.

**Synthesis of multilayer Ti$_3$C$_2$T$_x$ MXene**

According to the modified method by previous method[43,44], 2 g of Ti$_3$AlC$_2$ powder was slowly added to 100 mL of solution of 10 wt% HF and 2 g LiCl salt, stirring continuously at 25 °C for 24 h. After that, the precipitate was washed twice with 6 M HCl and degassed deionized (DI) water. Finally, the collected product was dried under a vacuum at 80 °C for 4 h to obtain multilayer Ti$_3$C$_2$T$_x$ MXene (*m*-MXene) powder.

**Synthesis of multilayer Ti$_2$CT$_x$ MXene**

0.5 g Ti$_2$AlC powder was slowly immersed into 40 mL 49% HF solution and continuously reacted at 25 °C for 10 h[45]. After washing with DI water until pH up to ~6, the product was filtrated and dried under vacuum at 80 °C for 4 h to obtain multilayer Ti$_2$CT$_x$ MXene powder.

**Synthesis of multilayer Nb$_2$CT$_x$ MXene**

1 g of Nb$_2$AlC powder was added to 20 mL of a solution of 49 wt% HF and 1 g LiCl salt, followed by stirring continuously at 55 °C for 28 h[25]. After that, the mixture underwent centrifugation to separate the powder from the solution, and the resulting precipitate was washed twice with 6 M HCl and degassed deionized (DI) water.

**Synthesis of multilayer TiNbCT$_x$ MXene**

0.5 g of TiNbAlC powder was slowly immersed into 40 mL of 49% HF solution and continuously reacted at 25 °C for 19 h[43]. After washing with DI water until pH of ~6, the product was filtrated and dried under vacuum at 80 °C for 4 h to obtain multilayer TiNbCT$_x$ MXene powder for electrode materials.

**Synthesis of multilayer Ti$_2$Mo$_2$CT$_x$ MXene**

1 g Ti$_2$Mo$_2$AlC$_3$ was slowly added into 20 mL of 49% HF solution and etched at 55 °C for 96 h[43]. Then, the obtained product was washed adequately until the pH of the last centrifugal supernatant reached around 6.0. Finally, the sediment was dried under vacuum at 80 °C for 4 h to obtain multilayer Ti$_2$Mo$_2$CT$_x$ MXene powder for electrode materials.

**Synthesis of multilayer Ti$_3$CNT$_x$ MXene**

1 g of Ti$_3$AlCN powder was added into 10 ml of 30 % HF solution and etched at 25 °C for 18 h[46]. After approximately six washes with deionized water until the pH reached ~6, the product was filtrated and dried under vacuum at 80 °C for 4 h to obtain multilayer Ti$_3$CNT$_x$ MXene powder.

**Synthesis of multilayer Ti$_3$C$_2$Cl$_2$ or Ti$_3$C$_2$Br$_2$ MXenes**

To produce Ti$_3$C$_2$Cl$_2$, 0.3 g of Ti$_3$AlC$_2$ powder was mixed with CdCl$_2$, NaCl, and KCl salts at the ratio of 1:3:6:6 by using a mortar and pestle under argon protection in a glove box[24]. The resulting mixture was removed from the glove box and heated in an alumina crucible under Ar at 650 °C for 5 h with a ramping rate of 5 °C/min. To produce Ti$_3$C$_2$Br$_2$, 0.2 g of Ti$_3$AlC$_2$ powder was mixed with CdBr$_2$ at the ratio of 1:13. The resultant mixture was heated in an alumina crucible at 680 °C for at least 12 h[24]. After washing with HCl or HBr solutions and DI water, the products were filtrated and dried under vacuum at 80 °C for 4 h to obtain Ti$_3$C$_2$Cl$_2$ and Ti$_3$C$_2$Br$_2$ powders.

**Preparation of organometallic-inorganic MXene or TMD hybrids**

About 0.1 g of *m*-MXene or transition metal dichalcogenide (TMDs) powders were compressed into a cylindrical pellet (Φ14 mm) under the pressure of 2 MPa. During a typical electrochemical intercalation process, the pellet was located at the bottom of a graphite crucible as the cathode, while a graphite rod served as the anode. Different organometallic compounds, including [Co(Cp)$_2$]PF$_6$, FcNCl, CoPc, [Ru(NH$_2$)$_6$]Cl$_3$ or [Pt(NH$_2$)$_4$](NO$_3$)$_2$ were used as electrolytes. The electrolyte solutions were prepared by dissolving 0.1 g of an organometallic compound in 5 g of DMSO, PC, or ACN solvents. A constant current was applied to the device to facilitate the migration of cations into the cathode over various time intervals to achieve uniform intercalation. For the contrast experiment to explore the influence of co-intercalated solvents or intercalating FcNCl species, the current was set to 0.5 mA owing to the limitations of work electrodes, and all others were intercalated under 2 mA. At the end of the intercalation reaction, the products were washed several times with DI water to eliminate any residues before being vacuum-dried at 80 °C for further analysis and characterization.

**Characterization**

Structural characterization was performed using a Bruker D8 ADVANCE X-ray diffractometer with Cu K$_α$ radiation (λ=0.154 nm) to obtain X-ray diffraction (XRD) patterns. The morphologies and components of synthesized organometallic-inorganic MXene hybrids were characterized by a thermal field emission scanning electron microscope (SEM, Thermo Scientific, Verios G4 UC, USA) equipped with an energy dispersive X-ray spectroscopy (EDX) system. The atomic structural analyses were conducted using a FEI Spectra 300 TEM/STEM equipped with a double aberration corrector for high angle annular dark field (HAADF), electron energy loss spectroscopy (EELS), and energy dispersive spectrometer (EDS) characterizations. Various sample foils were prepared by a Helios-G4-CX focused ion beam (FIB) before TEM/STEM observation. Further characterization of strain distribution in Co/Fe-$m$-MXene samples was carried out using high-resolution transmission electron microscopy (HR-TEM) and scanning transmission electron microscopy (STEM) mode of the Talos F200x TEM. The functional groups in the compounds were confirmed by using Fourier transform infrared spectroscopy (FTIR, Bruker INVENIOR, Germany).

X-ray photoelectron spectroscopy (XPS) data was collected using a Shimadzu AXIS SUPRA+ with a monochromatized Al K$_\alpha$ X-ray source (1486.71 eV) to analyze elemental valence and chemical composition of as-prepared samples. The surfaces of the samples were cleaned by Ar$^+$ sputtering for 5 min before XPS characterizations. Raman scattering measurements were taken by the system (inVia-reflex, Renishaw, UK) with a 532 nm excitation wavelength operating at 0.5 % of its powder, and the samples were placed on a glass slide to be captured by a 50x magnifying lens. The changes in height and Young's modulus were measured by atomic force microscopy (AFM, Dimension Icon, Bruker). Ti K-edge X-ray absorption (XAS) characterizations were made in the transmission mode at the beamline 1W1B at Beijing Synchrotron Radiation Facility (BSRF). Ti L-edge XAS and resonant inelastic X-ray scattering (RIXS) measurements were carried out at the beamline 8.0.1.1 of Advanced Light Source (Lawrence Berkeley National Laboratory). The multi-channel Biologic VMP 3e potentiostat was utilized to apply constant current during the electrochemical intercalation and record I-t curves to monitor the process. CV curves were measured to further unveil the reaction mechanism of the organometallic intercalants.

The temperature-dependent resistivity measurements were conducted using a physical property measurement system (Quantum Design, PPMS-9). Concretely, the dried MXene powders were compressed into square pellets under a load of 150 MPa, with a length of 5 mm and a thickness of ~0.7 mm. Four copper-plated spring-loaded electrodes were used to electrically contact the MXene pellet to a PPMS puck for the standard 4-probe resistivity measurements. The temperature-dependent resistivity measurements were performed from 300 K to 2 K at different applied magnetic fields (0 to 8 T). The magnetic susceptibility measurement was carried out using a SQUID magnetometer (Quantum Design, MPMS SQUID XL 5). From the zero-field-cooled curve of Co-$m$-MXene, the magnetic susceptibility at 2 K was -0.01912 emu/(g*Oe) (Fig. 3D). Based on the characterization results including lattice constants $a$ (3.03 Å) and $c$ (31.26 Å), and the molecular mass, M (200.561 g/mol), the crystallographic density was calculated to be 2.68 g/cm$^3$. It is worth noting that the crystallographic density significantly decreased after intercalating cobaltocene species, mainly ascribed to enlarged interlayer spacing. And the superconducting volume fraction was estimated as 0.01912*2.68*4π*100 % = 64.4 %[24].

## References in methods

## Acknowledgments

We thank R.P. Cui and Prof. C.L. Wan from Tsinghua University for XRD data analysis and Prof. B. Zhao from Fudan University for GPA analysis. K.L. gratefully acknowledges financial support from Anglo American Resources Trading (China) Co., Ltd. This work was supported by National Natural Science Foundation of China (U23A2093, 12375279, and 12374035), Ten-Thousand Talents Plan of Zhejiang Province (No. 2022R51007), Ningbo Top-talent Team Program, and Youth Science and Technology Innovation Leading Talent Project of Ningbo (2024QL022). Work at the Molecular Foundry and Advanced Light Source was supported by the Office of Science, Office of Basic Energy Sciences, of the U.S. Department of Energy under Contract No. DE-AC02-05CH11231.

## Competing interests

K.L. and Q.F. are inventors on a patent application (File No. 202411735172.3) submitted by the Ningbo Institute of Materials Technology and Engineering, which covers the electrochemical intercalation protocol for constructing organometallic MXene superconductors.

## Data availability

The data that support the findings of this study are available from the corresponding author upon reasonable request. Source Data are provided with this paper.

## Supplementary Materials

This file contains Supplementary Methods, Supplementary Figs. 1-31, Supplementary Tables 1-5 and Supplementary References.